\input epsf
%%%%%%%%%%%%%%%%%%%%%%%%%%%%%%%%%%%%%%%%%%%%%%%%%%%%%%%%%%%%%%%%%
%                                                               %
%       FONT FAMILIES:                                          %
%                                                               %
%%%%%%%%%%%%%%%%%%%%%%%%%%%%%%%%%%%%%%%%%%%%%%%%%%%%%%%%%%%%%%%%%
%                                                               %
%       Define script letters as rsfs                           %
%               (or redefine as cal)                            %
%                                                               %
%                                                               %
%%%%%%%%%%%%%%%%%%%%%%%%%%%%%%%%%%%%%%%%%%%%%%%%%%%%%%%%%%%%%%%%%
\newfam\scrfam
\batchmode\font\tenscr=rsfs10 \errorstopmode
\ifx\tenscr\nullfont
        \message{rsfs script font not available. Replacing with calligraphic.}
        \def\scr{\cal}

\else   
        \font\sevenscr=rsfs7
        \font\fivescr=rsfs5
        \skewchar\tenscr='177 \skewchar\sevenscr='177 \skewchar\fivescr='177
        \textfont\scrfam=\tenscr \scriptfont\scrfam=\sevenscr
        \scriptscriptfont\scrfam=\fivescr
        \def\scr{\fam\scrfam}
        \def\cal{\scr}
\fi
%%%%%%%%%%%%%%%%%%%%%%%%%%%%%%%%%%%%%%%%%%%%%%%%%%%%%%%%%%%%%%%%%
%                                                               %
%       Blackboard bold (or redefine as boldface)               %
%                                                               %
%%%%%%%%%%%%%%%%%%%%%%%%%%%%%%%%%%%%%%%%%%%%%%%%%%%%%%%%%%%%%%%%%
\newfam\msbfam
\batchmode\font\twelvemsb=msbm10 scaled\magstep1 \errorstopmode
\ifx\twelvemsb\nullfont\def\Bbb{\bf}

        \message{Blackboard bold not available. Replacing with boldface.}
\else   \catcode`\@=11
        \font\tenmsb=msbm10 \font\sevenmsb=msbm7 \font\fivemsb=msbm5
        \textfont\msbfam=\tenmsb
        \scriptfont\msbfam=\sevenmsb \scriptscriptfont\msbfam=\fivemsb
        \def\Bbb{\relax\expandafter\Bbb@}
        \def\Bbb@#1{{\Bbb@@{#1}}}
        \def\Bbb@@#1{\fam\msbfam\relax#1}
        \catcode`\@=\active

\fi
%%%%%%%%%%%%%%%%%%%%%%%%%%%%%%%%%%%%%%%%%%%%%%%%%%%%%%%%%%%%%%%%%
%                                                               %
%       MORE FONTS:                                             %
%                                                               %
%%%%%%%%%%%%%%%%%%%%%%%%%%%%%%%%%%%%%%%%%%%%%%%%%%%%%%%%%%%%%%%%%
        \font\eightrm=cmr8              \def\xrm{\eightrm}
        \font\eightbf=cmbx8             \def\xbf{\eightbf}
        \font\eightit=cmti10 at 8pt     \def\xit{\eightit}
%%%     \font\eightit=cmti8             \def\xit{\eightit}
        \font\eighttt=cmtt8             \def\xtt{\eighttt}
        \font\eightcp=cmcsc8
        \font\eighti=cmmi8              \def\xold{\eighti}
        \font\ninerm=cmr9
	\font\nineit=cmti9
	\font\ninett=cmtt9
	\font\teni=cmmi10               \def\old{\teni}

        \font\twelvecp=cmcsc10 scaled\magstep1
        
        \font\fiverm=cmr5

\def\noblackbox{\overfullrule=0pt}
\noblackbox

%%%%%%%%%%%%%%%%%%%%%%%%%%%%%%%%%%%%%%%%%%%%%%%%%%%%%%%%%%%%%%%%%
%                                                               %
%       HEADLINE:                                               %
%                                                               %
%%%%%%%%%%%%%%%%%%%%%%%%%%%%%%%%%%%%%%%%%%%%%%%%%%%%%%%%%%%%%%%%%
\headline={\ifnum\pageno=1
\else
{\eightcp Martin Cederwall:
        ``World-volume fields and background coupling of branes''}
                \dotfill{ }{\old\folio}\fi}
\def\makeheadline{\vbox to 0pt{\vss\noindent\the\headline\break
\hbox to\hsize{\hfill}}
        \vskip2\baselineskip}
%%%%%%%%%%%%%%%%%%%%%%%%%%%%%%%%%%%%%%%%%%%%%%%%%%%%%%%%%%%%%%%%%
%                                                               %
%       FOOTNOTES:                                              %
%                                                               %
%%%%%%%%%%%%%%%%%%%%%%%%%%%%%%%%%%%%%%%%%%%%%%%%%%%%%%%%%%%%%%%%%
%\footline={\ifnum\pageno=1\else\hfil\folio\hfil\fi}
\def\makefootline{
        \ifnum\foottest>0
                \ifnum\foottest=1
                        \footline={\footlineone}
                \fi
                \ifnum\foottest=2
                        \footline={\footlineone\footlinetwo}
                \fi
                \ifnum\foottest=3
                        \footline={\footlineone\footlinetwo\footlinethree}
                \fi
                \baselineskip=.8cm\vtop{\the\footline}
                \global\foottest=0
        \fi
        }
\newcount\foottest
\foottest=0
\def\footnote#1#2{${}^#1$\hskip-3pt
        \ifnum\foottest=2
        \def\footlinethree{\hfill\break
        \vtop{\baselineskip=9pt
        \indent ${}^#1$ \vtop{\hsize=14cm\noindent\xrm #2}}}\foottest=3
        \fi
        \ifnum\foottest=1
        \def\footlinetwo{\hfill\break
        \vtop{\baselineskip=9pt
        \indent ${}^#1$ \vtop{\hsize=14cm\noindent\xrm #2}}}\foottest=2
        \fi
        \ifnum\foottest=0
        \def\footlineone{\vtop{\baselineskip=9pt
        \hrule width.6\hsize\hfill\break
        \indent ${}^#1$ \vtop{\hsize=14cm\noindent\xrm #2}}
        \vskip-.7\baselineskip}
        \foottest=1
        \fi
        }
%%%%%%%%%%%%%%%%%%%%%%%%%%%%%%%%%%%%%%%%%%%%%%%%%%%%%%%%%%%%%%%%%
%                                                               %
%       REFERENCES:                                             %
%                                                               %
%%%%%%%%%%%%%%%%%%%%%%%%%%%%%%%%%%%%%%%%%%%%%%%%%%%%%%%%%%%%%%%%%
\newcount\refcount
\refcount=0
\newwrite\refwrite
\def\ref#1#2{\global\advance\refcount by 1
        \xdef#1{{\old\the\refcount}}
        \ifnum\the\refcount=1
        \immediate\openout\refwrite=\jobname.refs
        \fi
        \immediate\write\refwrite
                {\item{[{\xold\the\refcount}]} #2\hfill\par\vskip-2pt}}
\def\xref#1#2#3{\global\advance\refcount by 1
        \xdef#1{{\old\the\refcount}}
	\xdef#2{{\xold\the\refcount}}
	\ifnum\the\refcount=1
        	\immediate\openout\refwrite=\jobname.refs
        \fi
        \immediate\write\refwrite
                {\item{[{\xold\the\refcount}]} #3\hfill\par\vskip-2pt}}
\def\refout{\catcode`\@=11
        \xrm\immediate\closeout\refwrite
        \vskip2\baselineskip
        {\noindent\twelvecp References}\hfill\vskip\baselineskip
                                                %\vskip.25\baselineskip%%%%
        %\parskip=.875\parskip
        %\baselineskip=.8\baselineskip
        \baselineskip=.75\baselineskip
        \input\jobname.refs
        %\parskip=8\parskip \divide\parskip by 7
        %\baselineskip=1.25\baselineskip
        \baselineskip=4\baselineskip \divide\baselineskip by 3
        \catcode`\@=\active\rm}
%%%%%%%%%%%%%%%%%%%%%%%%%%%%%%%%%%%%%%%%%%%%%%%%%%%%%%%%%%%%%%%%%
%                                                               %
%       SECTION NUMBERING:                                      %
%                                                               %
%%%%%%%%%%%%%%%%%%%%%%%%%%%%%%%%%%%%%%%%%%%%%%%%%%%%%%%%%%%%%%%%%
\newcount\sectioncount
\sectioncount=0
\def\section#1#2{\global\eqcount=0
        \global\advance\sectioncount by 1
        \vskip\baselineskip\noindent
        \hbox{\twelvecp\the\sectioncount. #2\hfill}\vskip\baselineskip\noindent
        \xdef#1{{\old\the\sectioncount}}}
\newcount\appendixcount
\appendixcount=0
\def\appendix#1{\global\eqcount=0
        \global\advance\appendixcount by 1
        \vskip2\baselineskip\noindent
        \ifnum\the\appendixcount=1
        \hbox{\twelvecp Appendix A: #1\hfill}\vskip\baselineskip\noindent\fi
    \ifnum\the\appendixcount=2
        \hbox{\twelvecp Appendix B: #1\hfill}\vskip\baselineskip\noindent\fi
    \ifnum\the\appendixcount=3
        \hbox{\twelvecp Appendix C: #1\hfill}\vskip\baselineskip\noindent\fi}

%%%%%%%%%%%%%%%%%%%%%%%%%%%%%%%%%%%%%%%%%%%%%%%%%%%%%%%%%%%%%%%%%
%                                                               %
%       EQUATION NUMBERING                                      %
%                                                               %
%%%%%%%%%%%%%%%%%%%%%%%%%%%%%%%%%%%%%%%%%%%%%%%%%%%%%%%%%%%%%%%%%
\newcount\eqcount
\eqcount=0
\def\Eqn#1{\global\advance\eqcount by 1
        \xdef#1{{\old\the\sectioncount}.{\old\the\eqcount}}
        \ifnum\the\appendixcount=0
                \eqno({\oldstyle\the\sectioncount}.{\oldstyle\the\eqcount})\fi
        \ifnum\the\appendixcount=1
                \eqno({\oldstyle A}.{\oldstyle\the\eqcount})\fi
        \ifnum\the\appendixcount=2
                \eqno({\oldstyle B}.{\oldstyle\the\eqcount})\fi
        \ifnum\the\appendixcount=3
                \eqno({\oldstyle C}.{\oldstyle\the\eqcount})\fi}
\def\eqn{\global\advance\eqcount by 1
        \ifnum\the\appendixcount=0
                \eqno({\oldstyle\the\sectioncount}.{\oldstyle\the\eqcount})\fi
        \ifnum\the\appendixcount=1
                \eqno({\oldstyle A}.{\oldstyle\the\eqcount})\fi
        \ifnum\the\appendixcount=2
                \eqno({\oldstyle B}.{\oldstyle\the\eqcount})\fi
        \ifnum\the\appendixcount=3
                \eqno({\oldstyle C}.{\oldstyle\the\eqcount})\fi}
\def\multi{\global\advance\eqcount by 1}
\def\multieq#1#2{\xdef#1{{\old\the\eqcount#2}}
        \eqno{({\oldstyle\the\eqcount#2})}}
%%%%%%%%%%%%%%%%%%%%%%%%%%%%%%%%%%%%%%%%%%%%%%%%%%%%%%%%%%%%%%%%%
%                                                               %
%       Hyperrefs:                                        	%
%                                                               %
%%%%%%%%%%%%%%%%%%%%%%%%%%%%%%%%%%%%%%%%%%%%%%%%%%%%%%%%%%%%%%%%%
\newtoks\url
\def\href#1#2{{#2}}
\def\Href#1#2{\catcode`\#=12\url={#1}\catcode`\#=\active#2}

\def\HHref#1{\catcode`\#=12\url={#1}\catcode`\#=\active#1}
%%%%%%%%%%%%%%%%%%%%%%%%%%%%%%%%%%%%%%%%%%%%%%%%%%%%%%%%%%%%%%%%%
%                                                               %
%       FORMAT:                                                 %
%                                                               %
%%%%%%%%%%%%%%%%%%%%%%%%%%%%%%%%%%%%%%%%%%%%%%%%%%%%%%%%%%%%%%%%%
\parskip=3.5pt plus .3pt minus .3pt
\baselineskip=14pt plus .05pt minus .05pt
\lineskip=.5pt plus .05pt minus .05pt
\lineskiplimit=.5pt
\abovedisplayskip=18pt plus 2pt minus 2pt
\belowdisplayskip=\abovedisplayskip
\hsize=15cm
\vsize=19cm
\hoffset=.8cm
\voffset=1.8cm
\frenchspacing
%%%%%%%%%%%%%%%%%%%%%%%%%%%%%%%%%%%%%%%%%%%%%%%%%%%%%%%%%%%%%%%%%
%                                                               %
%       VARIOUS DEFINITIONS                                     %
%                                                               %
%%%%%%%%%%%%%%%%%%%%%%%%%%%%%%%%%%%%%%%%%%%%%%%%%%%%%%%%%%%%%%%%%
\def\ss{\scriptstyle}

\def\/{\over}
\def\*{\partial}
\def\punkt{\,\,.}
\def\komma{\,\,,}
\def\minus{\!-\!}
\def\+{\!+\!}
\def\={\!=\!}
\def\small#1{{\hbox{$#1$}}}
\def\half{\small{1\/2}}

\def\eg{{\tenit e.g.}}
\def\ie{{\tenit i.e.}}

\def\a{\alpha}
\def\d{\delta}
\def\k{\kappa}
\def\L{\Lambda}
\def\S{\Sigma}
\def\M{{\cal M}}
\def\R{{\Bbb R}}
\def\Z{{\Bbb Z}}

\def\II{\hbox{I\hskip-0.6pt I}}

\def\w{\wedge}
\def\Int{\int\limits}

\def\nl{\hfill\break\indent}
\def\nlni{\hfill\break}

%
%
%
%	THE PAPER
%
%
\null\vskip-2cm
\vtop{\baselineskip=9pt
\line{\hfill{\xrm G\"oteborg-ITP-98-07}}
\line{\hfill{\xtt hep-th/9806151}}
\line{\hfill{\xrm June, 1998}}}

\vskip1cm
\centerline{\tenbf World-volume fields and background coupling of branes}
 
\vskip3\parskip
\centerline{\ninerm MARTIN CEDERWALL}

\vskip\parskip
\catcode`\@=11
\centerline{\nineit Institute for Theoretical Physics}\vskip-4pt
\centerline{\nineit G\"oteborg University 
		and Chalmers University of Technology }\vskip-4pt
\centerline{\nineit S-412 96 G\"oteborg, Sweden}
\centerline{\ninett tfemc@fy.chalmers.se}\vskip-4pt
\centerline{\ninett 
\Href{http://fy.chalmers.se/~tfemc/}{http://fy.chalmers.se/\~{}tfemc/}}
\catcode`\@=\active

\vskip\parskip
\noindent\hskip1.5cm\vtop{\hsize=12cm \baselineskip=9pt\xit \noindent
This is the written version of an invited talk delivered at the
workshop ``Quantum gravity in the Southern Cone'' held in San Carlos
de Bariloche, Argentina, January 7-10, 1998.
After giving a brief introduction to the concept of branes and their
r\^ole in string theory, this talk describes a method for formulating
the dynamics of branes, especially those containing non-scalar moduli.
Emphasis is put on the coupling of branes to fields in the low-energy
background supergravity theories, and on preservation of maximal
amount of manifest symmetry. Due to the nature of the workshop,
the presentation is aimed at physicists who are not experts in string 
theory.}

\def\hepth#1{\href{http://xxx.lanl.gov/abs/hep-th/#1}{{\xtt hep-th/#1}}}
\def\jhep#1#2#3{\href{http://jhep.sissa.it/stdsearch?paper=#1\%28#2\%29#3}{{\xit JHEP} {\xbf #1} ({\xold#2}) {\xold#3}}}

\ref\Review{C.M. Hull and P.K. Townsend,
        {\xit ``Unity of superstring dualities''},
        Nucl.Phys. {\xbf B438} ({\xold1995}) {\xold109}
	[\hepth{9410167}];
\nlni E. Witten, {\xit ``String theory dynamics in various dimensions''},
        Nucl.Phys. {\xbf B443} ({\xold1995}) {\xold85}
        [\hepth{9503124}];\nlni
        J.H. Schwarz, {\xit ``The power of M theory''},
        Phys.Lett. {\xbf B367} ({\xold1996}) {\xold97}
        [\hepth{9510086}];\nlni
        P.K. Townsend, {\xit ``Four lectures on M-theory''},
        \hepth{9612121}.}
\xref\Hull\xHull{C.M. Hull,
	{\xit ``Gravitational duality, branes and charges''},\nl
	Nucl. Phys. {\xbf B509} ({\xold1998}) {\xold216} 
	[\hepth{9705162}].}
\ref\Moduli{D.M. Kaplan and J. Michelson, 
	{\xit ``Zero modes for the D 11 membrane and five-brane''},\nl
	Phys. Rev. {\xbf D53} ({\xold1996}) {\xold3474}
	[\hepth{9510053}];\nlni
C. Callan Jr., J. Harvey and A. Strominger,
	{\xit ``World-brane actions for string solitons''},\nl
	Nucl. Phys. {\xbf B367} ({\xold1991}) {\xold60};\nlni
T. Adawi, M. Cederwall, U. Gran, B.E.W. Nilsson and B. Razaznejad,
	{\xit forthcoming publication}.}
\ref\Polchinski{J. Polchinski, {\xit ``Dirichlet-branes and Ramond--Ramond 
	charges''},
	\nl Phys.~Rev.~Lett.~{\xbf 75} ({\xold1995}) {\xold4724} 
	[\hepth{9510017}].}
\ref\Supergravities{
L.~Brink and P.~Howe, {\xit ``Eleven-dimensional supergravity 
	on the mass-shell in superspace''},
	\nl Phys.~Lett.~{\xbf 91B} ({\xold1980}) {\xold384};\nlni
E.~Cremmer and S.~Ferrara, 
	{\xit ``Formulation of eleven-dimensional supergravity 
	in superspace''},
	\nl Phys.~Lett.~{\xbf 91B} ({\xold1980}) {\xold61};\nlni
P.S.~Howe and P.C.~West, 
	{\xit ``The complete N=2, d=10 supergravity''},
	Nucl.~Phys.~{\xbf B238} ({\xold1984}) {\xold181};\nlni 
J.L.~Carr, S.J.~Gates Jr.~and R.N.~Oerter, 
	\nl{\xit ``D=10, N=2A supergravity in superspace''},
	Phys.~Lett.~{\xbf 189B} ({\xold1987}) {\xold68}.}
\ref\Leigh{R.G. Leigh, {\xit ``Dirac--Born--Infeld action from Dirichlet sigma
	model''}, Mod.~Phys.~Lett.~{\xbf A4} ({\xold1989}) {\xold2767};\nlni
	C.G. Callan, C. Lovelace, C.R. Nappi and S.A. Yost,
	{\xit ``String loop corrections to beta functions''},
	\nl Nucl.~Phys. {\xbf B288} ({\xold1987}) {\xold525}.}
\ref\Douglas{M.~Douglas, {\xit ``Branes within branes''},
	\hepth{9512077};\nlni
M.B.~Green, C.M.~Hull and P.K.~Townsend, \nl{\xit ``D-brane 
	Wess--Zumino actions, T-duality and the cosmological constant''},\nl
	Phys. Lett. {\xbf B382} ({\xold1996}) {\xold65} 
	[\hepth{9604119}].}
\ref\SuperDbranes{
	M.~Cederwall, A.~von~Gussich, B.E.W.~Nilsson and A.~Westerberg,\nl
	{\xit ``The Dirichlet super-three-brane in type IIB supergravity''},
	\nl Nucl. Phys. {\xbf B490} ({\xold 1997}) {\xold 163}
        [\hepth{9610148}];\nlni
M. Cederwall, A. von Gussich, B.E.W. Nilsson, P. Sundell
        and A. Westerberg, \nl{\xit ``The Dirichlet super-p-branes in
        type \II A and \II B supergravity''}, 
	\nl Nucl. Phys. {\xbf B490} ({\xold 1997}) {\xold 179}
        [\hepth{9611159}];\nlni
M. Aganagic, C. Popescu and J.H. Schwarz, 
	{\xit ``D-brane actions with local kappa symmetry''}, 
	\nl Phys. Lett. {\xbf B393} ({\xold 1997}) {\xold 311}
	[\hepth{9610249}];\nl
        {\xit ``Gauge-invariant and gauge-fixed D-brane actions''},
        Nucl. Phys. {\xbf B495} ({\xold 1997}) {\xold 99}
	[\hepth{9612080}];\nlni
E. Bergshoeff and P.K. Townsend, {\xit ``Super D-branes''}, 
	Nucl. Phys. {\xbf B490} ({\xold 1997}) {\xold 145} 
	[\hepth{9611173}].}
\ref\WittenDbranes{E. Witten, {\xit ``Bound states of strings and p-branes''},
        Nucl. Phys. {\xbf B460} ({\xold 1996})  {\xold 335}
	[\hepth{9510135}].}
\ref\CT{P.K. Townsend, {\xit ``Membrane tension and manifest
        \II B S-duality''}, Phys. Lett. {\xbf B409} ({\xold1997}) {\xold131} 
	[\hepth{9705160}];\nlni
M. Cederwall and P.K. Townsend, {\xit ``The manifestly 
	{\xrm SL(2;Z)}-covariant superstring''},\nl
	\jhep{09}{1997}{003} 
	[\hepth{9709002}].}
\ref\SLSchwarz{J.H. Schwarz, 
	{\xit ``An {\xrm SL(2;Z)} multiplet of 
	type \II B superstrings''}, 
	Phys. Lett. {\xbf B360} ({\xold 1995}) {\xold 13}
	\nl[\hepth{9508143}];
        Erratum: ibid. {\xbf B364} ({\xold 1995}) {\xold 252}.} 
\ref\TGG{A.A.~Tseytlin, {\xit ``Self-duality of Born--Infeld action and 
	Dirichlet 3-brane of Type IIB superstring''},
	\nl Nucl.~Phys.~{\xbf B469} ({\xold1996}) {\xold51}
	[\hepth{9602064}];\nlni
	M.B.~Green and M.~Gutperle, {\xit ``Comments on 3-branes''},
	Phys.~Lett.~\xbf B377 \xrm ({\xold1996}) {\xold28}
	[\hepth{9602077}].}
\ref\CW{M. Cederwall and A. Westerberg, 
	``{\xit World-volume fields, SL(2;Z) and duality: 
		The type IIB 3-brane}'', 
	\nl\jhep{01}{1998}{004}
	[\hepth{9710007}].}
\ref\CNS{M. Cederwall, B.E.W. Nilsson and P. Sundell, 
	``{\xit An action for the 5-brane in D=11 supergravity}'', 
	\nl\jhep{04}{1998}{007} 
	[\hepth{9712059}].}
\ref\Intract{M. Aganagic, J. Park, C. Popescu and J.H. Schwarz,\nl
	{\xit ``Dual D-brane actions''},
	Nucl. Phys. {\xbf B496} ({\xold1997}) {\xold215} 
	[\hepth{9702133}].}

\vskip4\parskip\noindent

\section\Intro{Introduction}Extended supersymmetric objects arise as
solutions in the low-energy effective field theories of
superstring theory/M-theory. 
Some of them are naturally interpreted as solitons, and carry information
about the non-perturbative behaviour of the theory, while some bear signs 
of being elementary excitations (w.r.t. to some perturbative formulation). 
Most profoundly, the solitons of type \II\ string theory are realised
as D-branes, \ie, objects on which elementary strings may end.
Many aspects of the dynamics of these objects, which is the main
subject of this paper, are by now fairly well understood. 
Although a fundamental microscopic formulation of the theory is still 
missing, the evidence is overwhelming that the different perturbative
superstring theories, their low-energy supergravity theories and 11-dimensional
supergravity are different manifestations of one underlying non-perturbative
theory.

In this talk, there is not much time to discuss the r\^oles that the 
different elementary and solitonic objects have in the various duality
transformations between different perturbative theories. 
We refer to ref. [\Review]. Some will be mentioned
briefly, but the emphasis is put on the description of the dynamics
of branes occurring
in superstring theory/M-theory.

\section\ChargedBranes{Charged branes}The means to describe brane dynamics 
is to formulate actions, fulfilling certain requirements. In this section,
some things will be said about the basic ingredients in these actions.
The methods will later be refined.
We let the dimensionality 
of the brane world-volume be $p\+1$ and that of
space-time $D$. 

The first invariant considered as (a term in) the action for a brane
is the volume. It may be written
$$
\Int_{\S_{p+1}}d^{p+1}\xi\sqrt{-g}\komma\Eqn\Volume
$$
where $g_{ij}=\*_iX^m\*_jX^ng_{mn}$ is the induced metric on the world-volume.
For dimensionality reasons, it must be multiplied by some dimensionful
constant, the brane tension. There is also a possibility that some
scalar field (if the background theory contains one) will enter inside
the integral. This term is referred to as the Nambu--Goto action.

If the spectrum of the background supergravity contains
a $(p\+1)$-form potential $C_{(p+1)}$ with field strength\footnote*{This 
relation between potential and field strength will later, in some cases
contain additional terms; in such cases one has to be careful about what
exactly is meant by charge.}
$H_{(p+2)}=dC_{(p+1)}$, a $p$-brane can couple
electrically to $C$ via a term in the action\footnote\dagger{Of course, not
only electric coupling is formulated this way; magnetic coupling is
achieved as electric coupling to a dual field strength 
$\ss \tilde C_{(D-p-3)}$.}
$$
q\Int_{\S_{(p+1)}}C_{(p+1)}\komma\Eqn\WessZumino
$$
where $q$ is the charge of the brane. 
As long as the brane has no boundary, this term is clearly invariant under
gauge transformations $\d_\L C=d\L$.
Together with the kinetic term for
$C$ we can rewrite the relevant part of the action as an integral over
the entire space-time $\M_D$ as
$$
\Int_{\M_D}\left(\,\half H\w{*}H+q\hat n\w C\,\right)\komma\Eqn\Haction
$$
$\hat n$ being the $(D\minus p\minus 1)$-form whose components are those
of an epsilon tensor normal to the brane and which has $\d$-function support
on the world-volume.
The equation of motion for $C$ derived from eq. (\Haction)
is modified to include an electric source term (the exact sign here depends
on conventions and on the degrees of the forms)
and the brane charge is given as the integral of ${*}H$ over any
topological $(D\minus p\minus 2)$-sphere surrounding the brane
(see fig. {\old1}),
$$
\Int_{S_{D-p-2}}{*}H=\Int_{B_{D-p-1}}d{*}H=\Int_{B_{D-p-1}}q\hat n=q
	\punkt\eqn
$$
It is important to note that $q$ is not a ``charge density'' on the brane, but
an object intrinsically defined only for branes of dimension $p\+1$.

\noindent\hskip2cm\vtop{\hsize=9cm
\epsfxsize=4cm
\epsffile{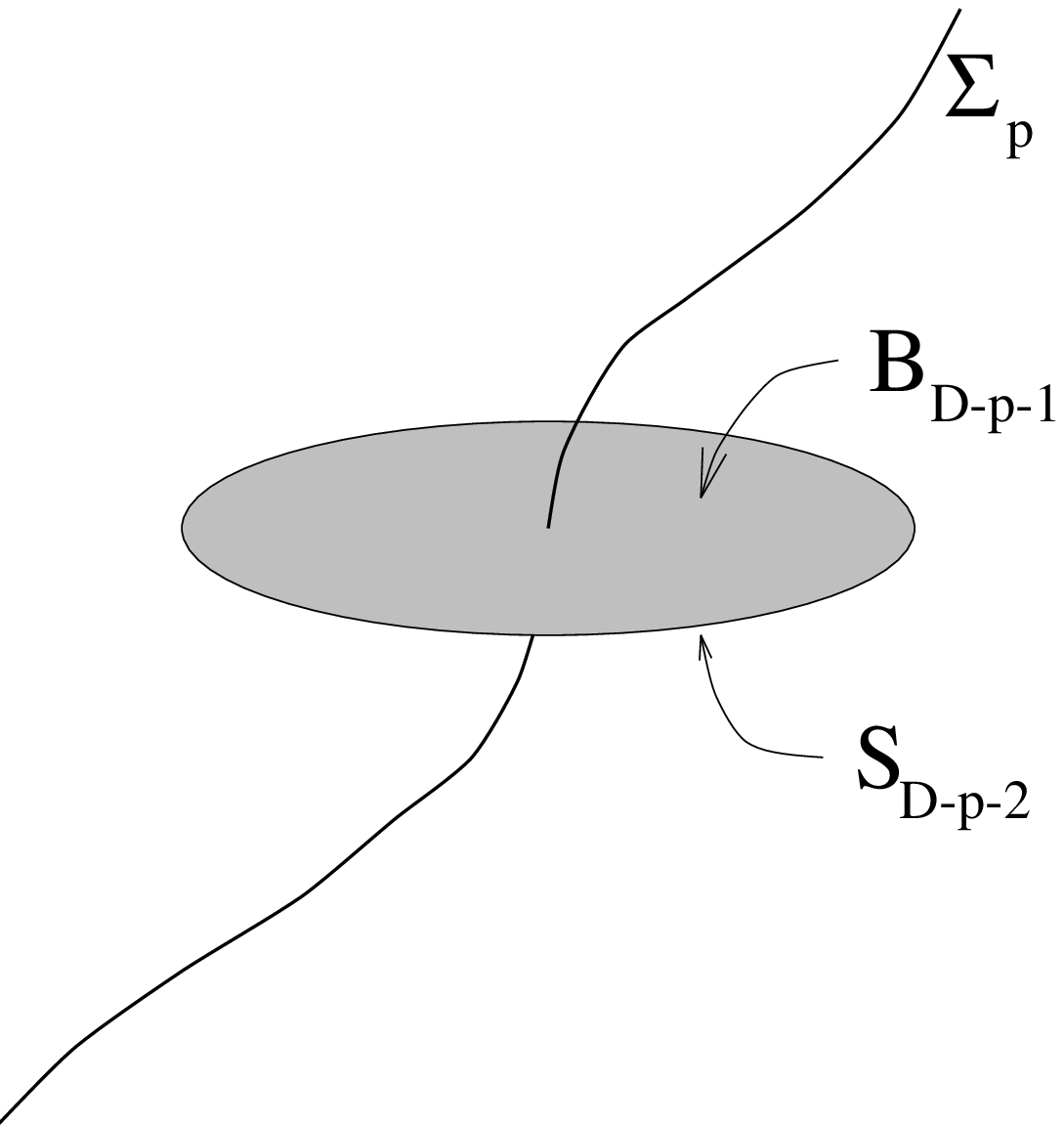}\hfill\break
{\it Figure 1. The topology in connection with brane charges. The time 
direction is not in the figure.}}
\vskip\parskip

From what has been said above,
it is clear that a charged $p$-brane may exist if the background supergravity
theory contains an antisymmetric tensor potential of rank $p+1$.
It is therefore sensible to look for solutions of the equations of motion
for the background supergravity fields that in some sense are localised
on a ``brane'', \ie, a lower-dimensional hypersurface, and where
the antisymmetric tensor carries some field strength corresponding to
a charge of that hypersurface.
It turns out that such solutions exist for all antisymmetric tensors
appearing in supergravity theories. For simplicity, we will concentrate
on maximally supersymmetric theories in maximum number of uncompactified
dimensions, which leaves us with type \II B supergravity
in 10 dimensions and 11-dimensional supergravity (type \II A in 10 dimensions
may be obtained as a dimensional reduction from 11 dimensions, and we
do not consider it separately here)\footnote*{It should be noted that
in addition to branes that carry tensorial charges, there are configurations
that are gravitationally charged (Kaluza-Klein monopoles, gravitational
waves [\xHull]). Although the discussion here is limited to
tensorial branes, compactification/dimensional reduction generically
mixes the two categories, since components of the metric appear as tensors
in the lower dimensionality.}.
Type \II B supergravity is the low-energy effective theory for type \II B
superstring theory, while 11-dimensional supergravity is related to
the so-called M-theory, whose fundamental degrees of freedom may reside
in the 11-dimensional supermembrane.

As an example of how brane solutions behave, we can take a look at
the tensorial branes of $D$=11 supergravity, a membrane and a 5-brane
(see following section).
The ``electric'' membrane solution contains a singularity, 
which is typical for an 
elementary excitation (such as an ordinary electrically charged
particle) --- it acts as a source . The ``magnetic'' 
5-brane solution, 
on the other hand, is non-singular
and in this sense exhibits a typically solitonic behaviour.
Even though solitonic solutions are non-singular, the charge is
protected by the topology of space-time.

Around brane solutions in supergravity, there will be zero-modes,
roughly speaking corresponding to flat directions in a moduli space
of such solutions. There has been surprisingly little work done on
such zero-modes and the derivation of brane dynamics from them [\Moduli], 
if one compares to the situation in field theory, where such methods are widely
used for \eg\ monopoles.

\vskip\parskip
\noindent\hskip2cm\vtop{\hsize=9cm
\epsfxsize=7cm
\epsffile{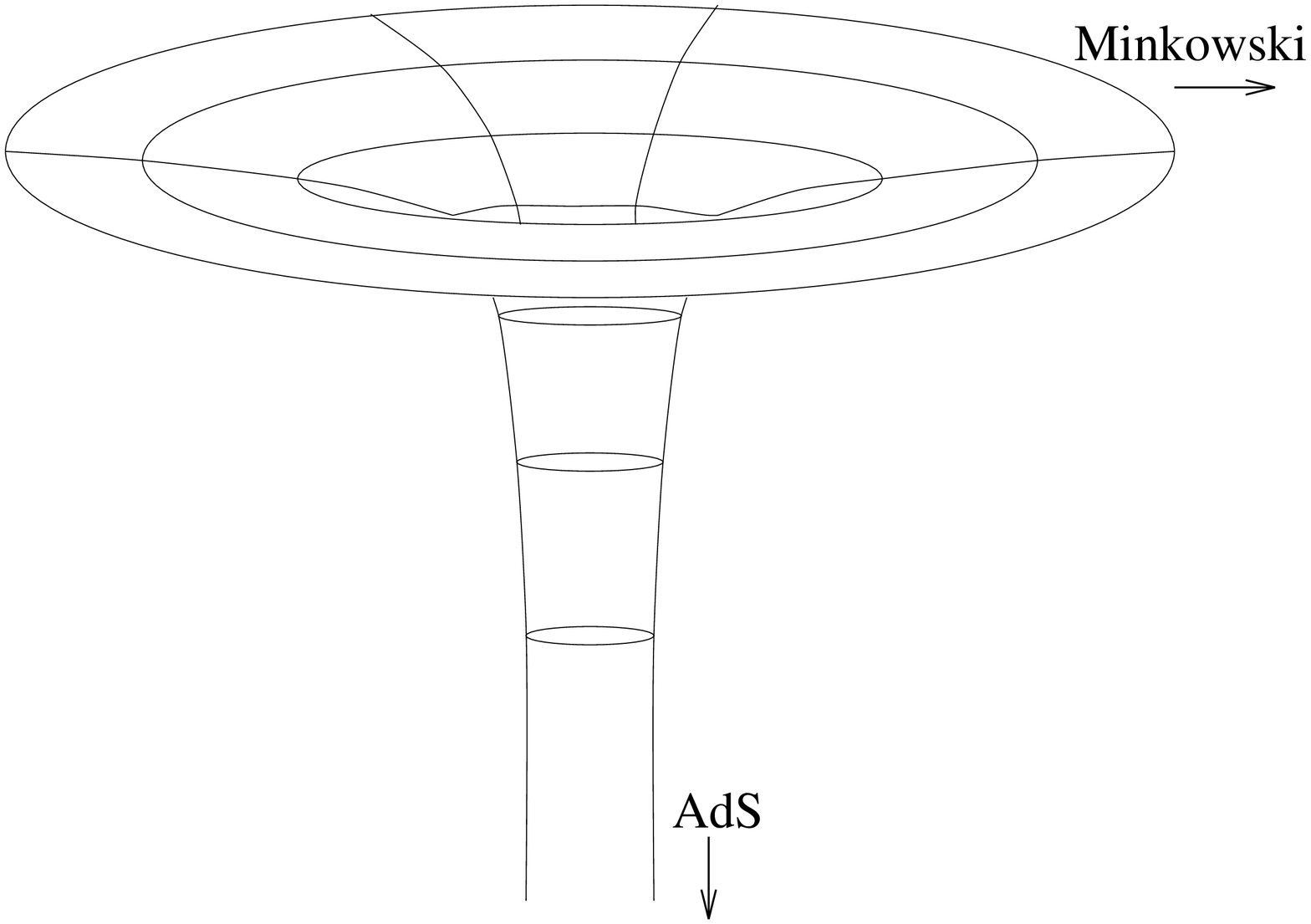}\hfill\break
{\it Figure 2. A rough picture of the space-like geometry associated with a 
brane solution in supergravity.}}
\vskip\parskip

An important property of the relevant branes is that they are stable.
Exactly as for solitons in supersymmetric field theory (monopoles...),
this is ensured by the BPS (Bogomol'nyi--Prasad--Sommerfeld) 
property, which in the simplest
case implies a relation between mass and charge, $m\sim q$. 
In supersymmetric theories, such a relation is exactly what is needed
for the existence of ``short multiplets'', smaller representations of the
supersymmetry algebra than the generic ones. The argument is roughly that
since the number of states in a multiplet should not get renormalised,
the BPS property is exact. BPS states are generated by a fraction of
the supersymmetry generators, which for single branes (the only
case treated here) is 1/2, but for multiple brane configurations may
be a smaller fraction.

The main topic of this talk is to describe the intrinsic dynamics
of the branes. They will be described by massless fields on the brane
world-volumes, belonging to supermultiplets generated by half the
number of space-time supersymmetry generators, due to the BPS property.

\section\GravityBranes{Supergravities and branes}We first list the
field content of the relevant supergravity theories. In type \II\
10-dimensional supergravities there are 128 bosonic and 128 fermionic
physical degrees of freedom (the numbers below refer to the dimensions
of the representations of the transverse rotation group Spin$(D-2)$) 
forming short ($2^8$) supermultiplets:
$$
\eqalign{
\left.\matrix{g&\hfill35\cr
	B_{(2)}/\tilde B_{(6)}&\hfill28\cr
	\phi&\hfill1\cr}
				\hskip3.5cm\right\}&\quad\hbox{NS-NS}\cr
}
$$
$$
\eqalign{
\matrix{\hbox to3.5cm{{\bf \II A}\hfill}
	&\hbox to3.5cm{\hfill{\bf \II B}\hfill}\cr}&\cr
\left.\matrix{
	\hbox to1.5cm{\hfill}
		&\hbox to1cm{\hfill}
		&\hbox to1cm{\hfill}
		&\hbox to2cm{\hfill}
		&\hbox to1cm{\hfill}\cr
	C_{(1)}/\tilde C_{(7)}&\hfill8&&
		C_{(0)}/\tilde C_{(8)}&\hfill1\cr
	C_{(3)}/\tilde C_{(5)}&\hfill56&&C_{(2)}/\tilde C_{(6)}&\hfill28\cr
	&&&C_{(4)}\hbox{ (chiral)}&\hfill35\cr}
				\qquad\right\}&\quad\hbox{RR}\cr
\left.\matrix{\hbox to1.5cm{\hfill}
		&\hbox to1cm{\hfill}
		&\hbox to1cm{\hfill}
		&\hbox to2cm{\hfill}
		&\hbox to1cm{\hfill}\cr
	\psi_\mu{}^\a&\hfill56&&\psi_{i\mu}{}^\a&\hfill2\times56\cr
	\psi_{\mu\a}&\hfill56&&&\cr
	\lambda_\a&\hfill8&&\lambda_{i\a}&\hfill2\times8\cr
	\lambda^\a&\hfill8&&&\cr}
				\qquad\right\}&\quad\hbox{NS-R, R-NS}\cr
}
$$
The fields in $D$=11 supergravity are
$$
\matrix{g&\hfill44\cr
	C_{(3)}/\tilde C_{(6)}&\hfill84\cr
	\psi_\mu{}^\a&\hfill128\cr}
$$
Here, $g$ denotes the metric, the $B$'s and $C$'s are antisymmetric tensor
potentials, and their dual potentials are denoted with a tilde. The
rank of the tensors is given in parentheses.  
The $\psi$'s are spin-3/2 gravitino fields and the $\lambda$'s spinors.
The letters NS or R denote the origin of the fields in perturbative
string theory, where they denote spin structures for left- and right-moving
fields on the string world-sheet.

From the table it is straightforward to read off the dimensionalities
of the different possible branes\footnote*{With tensorial charges; the
gravitational ones are again left out, resulting in an apparent 
mismatch between
type \II A and the reduction of $\ss D$=11.}:
$$
\matrix{\hbox{\bf \II B:}\hfill&\hbox to .5cm{\hfill}
	&p=(-1),1\hbox{ (2 charges)},3,5\hbox{ (2 charges)},7,(9)\hfill\cr
	\hbox{\bf \II A:}\hfill&
	&p=0,1,2,4,5,6,(8)\hfill\cr
	\hbox{\bf $\bf D$=11:}&
	&p=2,5\hfill\cr}
$$

We know that the degrees of freedom of the branes will form BPS multiplets
on the brane world-volumes. There are several ways of deducing exactly
what fields these multiplets consist of. Here we will make a very na\"\i ve
counting argument. First, count the number of fermionic degrees of freedom.
The background theory has (in all cases) 32 supercharges. Out of these,
only 16 generate physical states. On the brane, this BPS property manifests
itself as a fermionic gauge symmetry, ``$\k$-symmetry'', gauging
away half the spinor.
Then, accounting for the equation of motion for the world-volume fermions,
this number is further reduced to 8. Therefore, there must also be
8 bosonic degrees of freedom on the world-volume.
Some of these are obvious. There are translational degrees of freedom,
corresponding to moving the brane in a transverse direction. They
make up $D-p-1$, and there remains
$$
8-(D-p-1)=\left\{\matrix{p-1&(D=10)\komma\cr
			p-2&(D=11)\punkt\cr}\right.\eqn
$$
These numbers match with a world-volume U(1) vector potential 
($p-1$ degrees of freedom in $p+1$ dimensions) in $D$=10
and a chiral 2-form (3 degrees of freedom) 
for the 5-brane in $D$=11. It should perhaps be 
mentioned that this crude approach misses out
some interesting aspects, particularly in relation to the reduction
from 11 to 10 dimensions, but we try to keep it on a simple level.

\section\Actions{D-brane actions}We now turn to the issue of formulating
brane dynamics via an action principle. 
In the discussion above concerning the existence of tensorially charged
branes, no distinction was made between different ``types'' of branes.
From the point of view of a perturbative string theory, there is a clear
difference between fields in the NS-NS sector and in the RR sector.
A fundamental (perturbative) superstring couples only to NS-NS fields.
It was unclear for a long time how coupling to RR fields was to be achieved.
The solution, due to Polchinski [\Polchinski], is that the supergravity
solutions corresponding to RR-charged branes are non-perturbative ``states''
(using a somewhat sloppy language --- the solutions are of course classical,
and one should consider states in the corresponding quantum field theory)
in the superstring theory. As argued in the previous section, they have
massless excitations corresponding to a world-volume vector. This vector
couples to endpoints of open strings. The open strings are thus forced to 
end on the brane, \ie, to have Dirichlet boundary conditions.

\vskip\parskip
\noindent\hskip2cm\vtop{\hsize=9cm
\epsfxsize=5cm
\epsffile{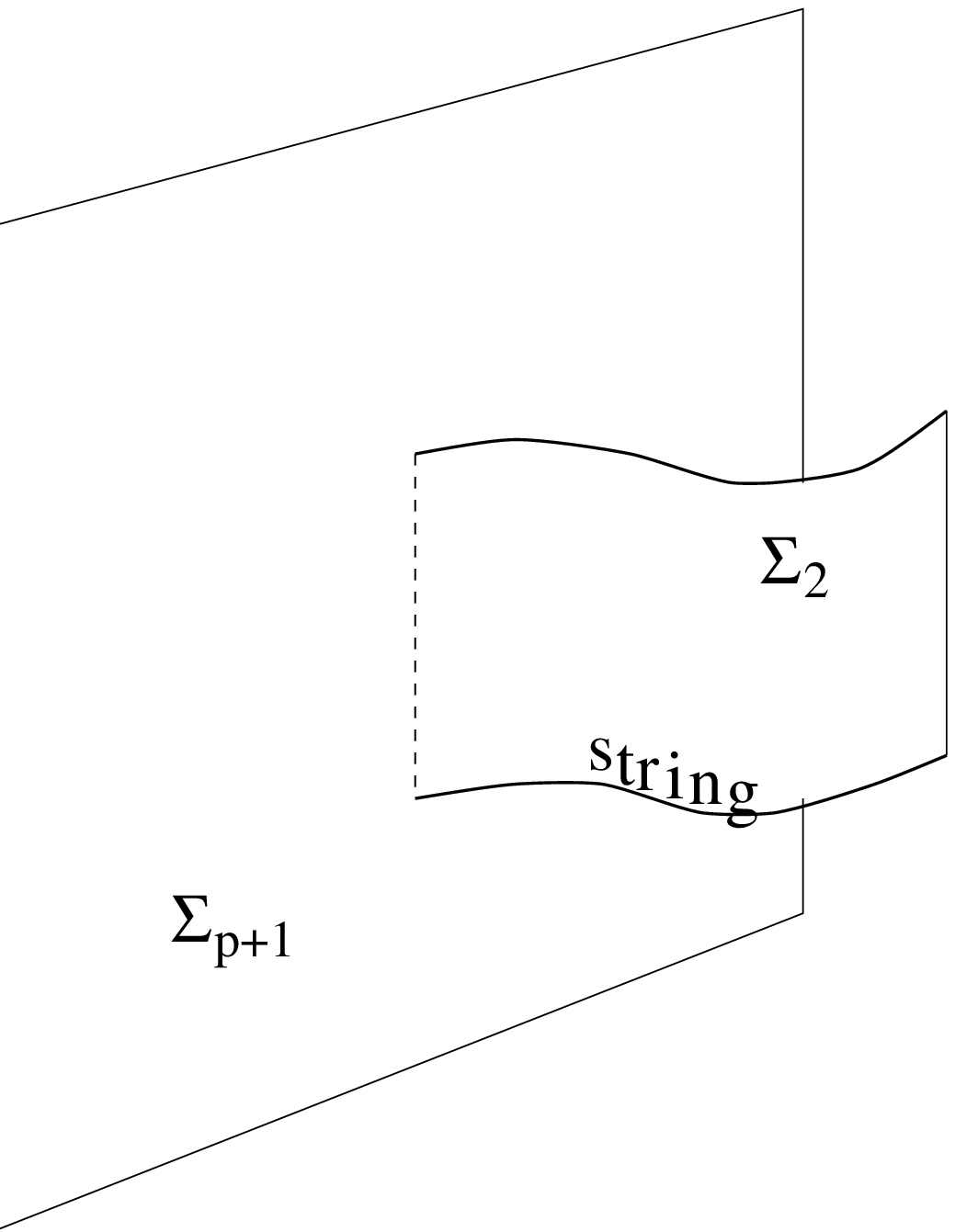}\hfill\break
{\it Figure 3. A string attached to a D-brane.}}
\vskip\parskip

Let us first consider actions for these D-branes, and later turn to
improvements of this description.
Some essential ingredients are present in section \Intro.
One piece in the action should be the invariant volume, as in eq. (\Volume),
and another one the tensor coupling of eq. (\WessZumino) (for simplicity,
tension factors are suppressed). The entire action should be some 
modification of the combination of these terms, also containing the
world-volume vector and maybe a dilaton coupling.

Some information about the way the vector enters may be obtained 
from gauge invariance in the configuration depicted in fig. {\old3}.
If we use the knowledge that the fundamental superstring couples
to the NS-NS 2-form potential $B$ with field strength $H=dB$, 
we can calculate the variation of
its action under a gauge transformation $\d B=d\Lambda$:
$$
\d\Int_{\S_2}B=\Int_{\S_2}d\Lambda=\Int_{\*\S_2}\Lambda\eqn
$$
($\*\S_2$ is the dotted line if fig. {\old3}).
This variation is cancelled if the string endpoint is considered as a
point source on the world-volume, so that there is a term $\int_{\*\S_2}A$,
provided that the world-volume vector potential $A$ transforms as
$\d A=\Lambda$ under these gauge transformations in the background theory.
The gauge invariant (both background and world-volume U(1)) field strength
of $A$ must then be $F=dA-B$, and this is the only combination through
which $A$ can enter the brane action.

As is known from type \II A/B supergravity, the RR tensor fields fulfill
Bianchi identities which are ``modified'' [\Supergravities].
If all the RR forms (odd for type \II A, even for type \II B) are collected
in $C=C_{(r)}\oplus C_{(r+2)}\oplus C_{(r+4)}\oplus\ldots$, where
$r=1$ for \II A and 0 for \II B, the field strengths are
$$
R=e^Bd(e^{-B}C)=dC-H\w C\punkt\eqn
$$
It is straightforward to verify that gauge invariance demands the Wess--Zumino
term to be modified as
$$
\Int_{\S_{p+1}}e^FC\punkt\eqn
$$

A $\beta$-function calculation [\Leigh] for the string ending on the
D-brane (one $\beta$-functional for each NS-NS background field, which
act as ``coupling constants'') shows that the $\beta$-functions may
be cancelled by the classical variation of
$$
-\Int_{\S_{p+1}}d^{p+1}\xi e^{-\phi}\sqrt{-\det(g+F)}\komma\eqn
$$
which is therefore considered as the appropriate modification of
the volume (``kinetic'') term in the D-brane action. This term is
known as the Dirac--Born--Infeld (DBI) action.

We have now, through different arguments, arrived at the final form
[\Douglas] of the action for the bosonic degrees of freedom of a D-brane:
$$
S=-\Int_{\S_{p+1}}d^{p+1}\xi e^{-\phi}\sqrt{-\det(g+F)}+\Int_{\S_{p+1}}e^FC
	\punkt\Eqn\DbraneAction
$$
How is this made supersymmetric?
The answer is straightforward: the bosonic world-volume is embedded in
superspace instead of in ordinary space. All background fields become
superfields, and enter the action (\DbraneAction), which is still
formally valid, as pullbacks from
superspace to the world-volume. Thus also the fermionic coordinates
in the background superspace become dynamical variables.
Their number is 32, and as mentioned earlier, there must be a fermionic
gauge symmetry to take away half of them, in accordance with the
BPS property of the brane. The symmetry is known as $\k$-symmetry,
and its construction and verification is a most central part of 
the formulation of the dynamics of a super-D-brane.
Let the superspace coordinates be $Z^M=(x^m,\theta^\mu)$, and the
locally inertial indices be $A=(a,\a)$. The $\k$-symmetry is most
covariantly formulated as a local fermionic translation
$\d_\k Z^M=\k^\a E_\a{}^M$, where $E_A{}^M$ is the super-vielbein.
If this symmetry is to remove only half a spinor, the spinor $\k$
must be subject to a relation $\k=P\k$, where $P$ is a projection
operator of half the maximal rank. We will not go through the details
of the construction, but refer to ref. [\SuperDbranes]. 
The projection operator may be written as
$P={1\/2}(1+\Gamma)$, where $\Gamma^2=1$, and takes the schematic form
(apart from numerical and dilaton factors)
$$
\Gamma d^{p+1}\xi\sim{1\/\sqrt{-\det(g+F)}}
	\left(\gamma_{(p-1)}+F\w\gamma_{(p+1)}
	+F\w F\w\gamma_{(p-3)}+\ldots\,\right)\komma\eqn
$$
the $\gamma_{(n)}$'s being $n$-forms whose components are antisymmetric
products of $\gamma$-matrices.

\section\TwoBString{More covariance in type \II B}In spite of the success
of the program presented in the previous section, some questions remain.
Some concern mainly type \II B, and some tensorial branes in general.

\vskip\parskip
\noindent\hbox to .5cm{$\bullet$\hfill}\vtop{\hsize=11cm 
\noindent Type \II B supergravity
has an SL(2;$\R$) invariance. The subgroup SL(2;$\Z$) is an S-duality
(strong-weak coupling symmetry) of type \II B superstring theory.
It is obscured by the (perturbative) division into NS-NS and RR sectors.
Can it be manifested?
}

\vskip1.5\parskip
\noindent\hbox to .5cm{$\bullet$\hfill}\vtop{\hsize=11cm 
\noindent What is special
about the $B$ field? Could not any $n$-form potential $C$ enter brane
actions through an $n$-form world-volume field strength
$F\sim dA-C$?
}
\vskip\parskip

The answer to the second question is in the affirmative. As we will
see, it also contains the answer to the first question.

The asymmetry pointed out here can be illustrated by the actions
for a fundamental string and a D-string in type \II B. They should be
related by transformations in SL(2;$\Z$).
The fundamental superstring is described by (from now on, all world-volumes
are considered to be embedded in superspace) 
$$
S=-\Int_{\S_2}d^2\xi\sqrt{-g}+\Int_{\S_2}B\komma\eqn
$$
while the D-string action is obtained from eq. (\DbraneAction) as
$$
S=-\Int_{\S_2}d^2\xi e^{-\phi}\sqrt{-\det(g+F)}
	+\Int_{\S_2}(C_{(2)}+FC_{(0)})\punkt\eqn
$$
The vector potential on the D-brane world-volume carries no local
degrees of freedom. There are however global ones, namely a quantised
electric flux [\WittenDbranes], giving charge with respect to $B$
(this is most easily seen in a canonical treatment). So while the
fundamental string has charges $\vec q=(1,0)$ w.r.t. $(B,C)$, the
D-brane action describes all the sectors with $\vec q=(m,1)$ for
integer $m$'s.

The asymmetry is cured by the introduction of yet another
world-volume vector potential. The two vectors now come in a doublet
of SL(2;$\R$), exactly as in the background supergravity.
One has to rescale the metric, since the one occurring so far was
the ``string metric'', whose curvature scalar is multiplied by a
dilaton factor in the supergravity action. The SL(2)-invariant metric
is the ``Einstein metric'', \ie, the one that comes without dilaton
factors in the Einstein--Hilbert action. They are related by
$$
g_{\hbox{\fiverm string}}=e^{\phi\/2}g_{\hbox{\fiverm Einstein}}\punkt
$$
The scalar fields $\phi$ and $C_{(0)}$ parametrise the coset
SL(2;$\R$)/U(1), and their covariance must also be taken care of.
The world-volume field strengths are most conveniently combined into
a complex 2-form $F$, which is invariant under $SL(2)$ and also contains 
the scalars. We refer to ref. [\CT] for details. The action turns
out to be astonishingly simple:
$$
S=-\Int_{\S_2}d^2\xi\lambda(1-{*}F{*}\bar F)\punkt\Eqn\SLAction
$$
We note that there is no Wess--Zumino term in this action, which is logical
since the tensors should not occur twice. The field $\lambda$ is a
Lagrange multiplier.
It is straightforward to
verify that eq. (\SLAction) gives the correct coupling to the background
fields and yields the correct equations of motion, \ie, equivalent to
the string actions above in those charge sectors, and that it enjoys
$\k$-symmetry. Due to flux
quantisation, the action describes
the entire set of strings with integer charges $\vec q=(m,n)$
(this is actually more than the orbit of $\vec q=(1,0)$ under SL(2;$\Z$)
consisting only of coprime pairs). For constant (or slowly varying)
scalar fields, one also verifies the correct string frame tension
[\SLSchwarz]
$$
T=\sqrt{e^{-2\phi}n^2+(m+nC_{(0)})^2}\punkt\eqn
$$

It is conceivable that the action (\SLAction) can be taken as a starting
point for the formulation of an SL(2;$\Z$)-covariant type \II B superstring
perturbation theory, where the quantised fluxes are conserved when
strings split or join. The situation seems much more hopeful than
for the S-duality of $N$=4 super-Yang--Mills theory, where duality mixes
electric and magnetic charges. This is not the case for type \II B, where
charges w.r.t. two different fields are mixed (the analogy of a magnetically
charged object is a 5-brane).

\section\HigherBranes{Higher-dimensional branes}How can the picture
of last section be generalised to other branes? The schematic ansatz
is that all coupling to background tensors should be achieved through
world-volume field strengths $F\sim dA-C$. The properties of the world-volume
fields should mimic the background ones, so that for type \II B one
would have a complex $F_{(2)}$, a real $F_{(4)}$, a complex $F_{(6)}$,
and so on. 

One problem presents itself immediately: while the
vector potential on the string world-sheet did not carry any local
degrees of freedom, these potential in general will, and one na\"\i vely
gets to many physical degrees of freedom. The solution is that there
will be an algebraic relation, a generalised self-duality, between
the field strengths. We will illustrate this first in the case of
the type \II B 3-brane.

The type \II B 3-brane is known to be ``self-dual'' [\TGG], in
the sense that a duality transformation of $F$ (which is necessarily 
non-linear, since the conjugate $E$-field obtained from the DBI action
is non-linear) accompanied by a $\Z_2$ in SL(2;$\Z$) leaves the
action invariant. What one now has to do is to find the appropriate
construction of the world-volume field strengths $F_{(2)}$ (complex) 
and $F_{(4)}$ (real). This is achieved through consideration of gauge
invariance. It is amusing to note that the modified Bianchi identities
of the background supergravity forces the world-volume Bianchi identities
to behave similarly (all details are found in ref. [\CW]).
With some guidance from the string case, a reasonable ansatz for an action
is
$$
S=\Int_{\S_4}d^4\xi\lambda
	\left(1+\Phi(F_{(2)},\bar F_{(2)})-({*}F_{(4)})^2\right)\komma\eqn
$$
with $\Phi$ being some yet unknown function. There are a number of
requirements, which all turn out to point to the same answer (in terms
of $\Phi$). One is that the exact form of the self-duality relation 
imposed on $F_{(2)}$ is to be consistent with the coupling to the
background fields; it is not a priori clear that equations of motion and
Bianchi identities contain identical background tensors. Another is that
the action should be $\k$-symmetric. Using this criterion leads to
the conclusion that $\k$-symmetry demands a specific form of the
self-duality, which is identical to the one obtained from the first
requirement. Very schematically, the form of the self-duality reads
$$
i{*}F_{(4)}{*}F_{(2)}\sim F_{(2)}+F_{(2)}{}^2\bar F_{(2)}\komma
\Eqn\NLselfd
$$
corresponding to a $\Phi$ with quadratic and quartic terms. To linear level,
eq. (\NLselfd) reduces to the ordinary complex self-duality, implying that
one real component of the complex $F$ is the dual of the other. At the price
of breaking the SL(2)-covariance, one real vector potential may be eliminated,
and one is back at the field content of the DBI action (if also the 
non-dynamical $F_{(4)}$ is solved for).

Partial results have been obtained for the multiplet of $(m,n)$ type \II B
5-branes, but the work is not finished. It involves an interesting
duality relation between a complex 2-form and a real 4-form, some aspects
of which were described in ref. [\CW]. There is a possibility that
the problem is algebraically intractable [\Intract].
The 5-brane in 11-dimensional supergravity has been described
in an analogous way [\CNS]. 

A drawback of these formulations is of course
that the self-duality does not follow from the action principle, but
has to be supplemented by hand. For the 11-dimensional 5-brane, there
is no action, since the multiplet contains a chiral 2-form; for the branes
in type \II B it is the price one has to pay in order to keep the
S-duality symmetry. An advantage is that the actions, as well as the
self-duality relations, become polynomial, instead of taking the
non-polynomial DBI form.
A great advantage of the formulation is that it provides a natural 
framework for studying
branes, other than strings, ending on branes and coupling via their
boundaries to the world-volume tensor potentials, analogously to
the way a string ends on a D-brane (it may be seen as surprising
that the formulation exists for the 5-brane in $D$=11, since 5-branes
do not end on 5-branes). This aspect has not been worked
out in detail, but looks promising.

\vskip\baselineskip
\noindent{\tenbf Acknowledgement}

\noindent Different parts of the work mentioned here was done 
in collaboration with 
Tom Adawi, Ulf Gran, Alexander von Gussich, Magnus Holm, Bengt E.W. Nilsson, 
Behrooz Razaznejad, Per Sundell, Paul K. Townsend and Anders Westerberg. 
I would also like to thank the organisers of the workshop ``Quantum gravity
in the Southern Cone'' in San Carlos de Bariloche, Jan. 1998, for
their hospitality and helpfulness during the stay in Argentina.

\refout
\end